\documentclass{elsarticle-mod}
\usepackage{geometry}                		
\usepackage{graphicx}				
\usepackage{amssymb}
\usepackage{natbib}
\usepackage{fixltx2e}
\usepackage{float}
\begin{document}

\begin{frontmatter}

\title{On the rotation rates and axis ratios of the smallest known
near-Earth asteroids---the archetypes of the Asteroid Redirect Mission targets}

\author[pa]{Patrick Hatch}
\ead{phatch@uwo.ca}
\author[pa,cpsx]{Paul A. Wiegert\corref{cor1}}
\ead{pwiegert@uwo.ca}
\address[pa]{Department of Physics and Astronomy, The University of Western Ontario, London, N6A 3K7 CANADA}
\address[cpsx]{Centre for Planetary Science and Exploration, The University of Western Ontario, London, N6A 3K7 CANADA}
\cortext[cor1]{Corresponding author. Tel: 1-519-661-2111x81327  Fax: 1-519-661-2033 Email: pwiegert@uwo.ca}

\begin{abstract}
NASA's Asteroid Redirect Mission (ARM) has been proposed with the aim
to capture a small asteroid a few meters in size and redirect it into
an orbit around the Moon. There it can be investigated at leisure by
astronauts aboard an Orion or other spacecraft. The target for the
mission has not yet been selected, and there are very few potential
targets currently known. Though sufficiently small near-Earth
asteroids (NEAs) are thought to be numerous, they are also difficult
to detect and characterize with current observational facilities.
Here we collect the most up-to-date information on near-Earth
asteroids in this size range to outline the state of understanding of
the properties of these small NEAs. Observational biases certainly mean that
our sample is not an ideal representation of the true population of
small NEAs. However our sample is representative of the eventual
target list for the ARM mission, which will be compiled under very
similar observational constraints unless dramatic changes are made to
the way near-Earth asteroids are searched for and studied.

We collect here information on 88 near-Earth asteroids with diameters
less than 60 meters and with high quality light curves.  We find that
the typical rotation period is 40 minutes. Relatively few axis ratios are
available for such small asteroids, so we also considered the 92 smallest
NEAs with known axis ratios. This sample includes asteroids with diameters
up to 300~m. The mean and median axis ratios were 1.43 and 1.29.

Rotation rates much faster than the spin barrier are seen, reaching
below 30 seconds, and implying that most of these bodies are
monoliths. Non-principal axis rotation is uncommon. Axial ratios often
reach values as high as two, though no undisputed results reach above
three. We find little correlation of axis ratio with size. The most
common spectral type in the sample of small NEAs is S-type ($> 90\%$),
with only a handful of C and X types known.

\begin{keyword}
asteroids, composition; asteroids, rotation; Asteroid Redirect Mission (ARM); Near-Earth asteroids (NEAs) 
\end{keyword}

\end{abstract}
\end{frontmatter}

\section{Introduction}

In a detailed study of a hypothetical mission to retrieve a small
asteroid and bring it to near-Earth space, the Keck Institute for
Space Studies (KISS) report \citep{broculfri12}\footnote{Available at
  {\it
    http://kiss.caltech.edu/study/asteroid/asteroid\_final\_report.pdf},
  retrieved 2014 Jul 24} concluded that ``one of the most challenging
aspects of the mission was the identification and characterization of
target NEAs suitable for capture and return'' (p.7). The report also
outlines three key mission drivers, one of which is ``the size and
mass of the target body'' (p. 28); the two others are the total
delta-$v$ required for capture and return, and the total flight time.

The design of the Asteroid Redirect Mission or a similar mission
depends significantly on the properties of the target, namely its
mass, size, density, internal cohesiveness, spin state, surface
roughness, presence/absence of regolith and so forth. In the ideal
case, mission planners will have complete information on the target's
characteristics before launch. However, the near-Earth asteroids in
the appropriate size range, which we will refer to as Very Small
Asteroids or VSAs, are particularly difficult to characterize. They
are faint and spend only a short time (typically days) within easy
reach of Earth-based telescopes when they are first discovered, often
not returning to the Earth's vicinity for several years.

Only relatively few of the already-known near-Earth asteroid
population make suitable targets, as most known NEAs are simply too
big. The discovery rate of suitable asteroids for the ARM was
estimated in the KISS report \nocite{broculfri12} (Table 2) to be five
asteroids per year if a low-cost ground-based telescopic campaign was
begun specifically to search for such asteroids; however, such a
dedicated program is not yet in place.  The total known sample of
potential targets as of June 2014 is only nine\footnote{NASA Announces
  Latest Progress in Hunt for Asteroids,
  http://www.jpl.nasa.gov/news/news.php?release=2014-195, retrieved
  2014 Nov 9} and what is known of their properties is scattered
throughout the literature and internet. By collecting information on
the smallest known NEAs, we hope to make the discussion of relevant
design issues simpler.

The heliocentric orbit of the asteroid can be relatively easily
determined, requiring only a handful of astrometric measurements from
short imaging exposures, and the orbit provides enough information for
the mission to be launched and to arrive at its destination
successfully. Not that a high-precision orbit can necessarily be
determined from the few-day apparition of a newly discovered small
asteroid, but orbits are typically easier to measure than the
asteroid's physical and internal properties and this may limit how
accurately the density, spin state, taxonomy, etc. of the target is
known before the mission proper is launched.  Though there is always
the option to study the target intensively when it makes a subsequent
passage near the Earth, these opportunities may occur only at
intervals of years, decades or even longer, and waiting for them could
delay the mission significantly.

Furthermore, even careful study may not reveal all the properties of
interest of a particular target. \citet{herwhi11} and
\citet{kwipolloa10} examined the light curves of many small asteroids,
and they point out that these are not always
conclusive. Non-detections of asteroid brightness variations could
indicate a non-rotating body, but could also be the result of asteroid
shapes that are close to spherical, viewed pole-on and/or with rapid
rotation periods that are not properly sampled by the exposure times
used. Since smaller asteroids have a tendency to rotate rather quickly
compared to large ones \citep{prahar00}, issues of this sort
complicate the picture. Studies such as the present paper of the
properties of the ARM target population as a whole can shed light on
the probable characteristics of individual targets for which some
properties cannot be measured prior to launch.

Considering as well the long flight time for the ARM (six to ten
years), it is conceivable that an incompletely characterized asteroid
with a particularly favorable orbit (i.e. one that would result in a
shorter travel time or a lower delta-$v$, and hence a lower cost for
the mission) might be more enticing as a candidate than a
better-studied small asteroid whose orbit is less favorable.  As a
result, a statistical study of the properties of small asteroids in
general provides helpful insight as to the likely or worst-case
properties of a potential target that is not yet fully characterized.

In the following sections, we collect the information available on VSAs in an
attempt to paint a picture of a typical asteroid within the size range
suitable to be an ARM target. This picture will include the most
likely spin-state, shape, and composition of such an asteroid. In
addition, we will also discuss the ``worst-case'' scenario for an ARM
target in terms of extremes of rotation rate and the likelihood of a tumbler or
non-principal axis (NPA) rotator.

\section{Methods}

The body of results on asteroids within the desired size range is
small. A large portion of the information presented here was gathered from
the Light Curve Database \citep[LCDB,][]{warharpra09}. Additional information
was collected from published asteroid surveys presented by
\citet{whithoher02}, \citet{kwibucodo10}, \citet{kwipolloa10},
\citet{herwhi11}, \citet{herkwikry12}, \citet{polbinloc12}
and \citet{stacotrie13}.

In obtaining data from the LCDB and the other surveys, we selected two
samples. One contained the smallest asteroids with known rotation periods,
and one the smallest asteroids with known axis ratios, as unfortunately not
all small asteroids have measurements of both of these quantities.

The first sample was selected on two criteria. Firstly, given the
scarcity of data on asteroids with diameters of ten meters and under
we chose a sample of asteroids with estimated diameters of 60 meters
and under as a proxy. The choice of 60 meters as our upper-boundary is
arbitrary, but it gave us a sizable amount of data without straying
too far from the intended diameter.  It also allows some consideration
of the alternative ARM scenario nicknamed {\it Pick Up A Rock}, where
instead of retrieving an asteroid whole (the {\it Get a Whole One}
scenario), a boulder or other material would be recovered from the
surface of a larger body. Secondly, for data that came directly from
the LCDB, asteroids with a quality rating $U$ lower than $2-$ were not
included in the study (The LCDB quality rating runs from 1 (low)
through 1+, 2-, 2, 2+, 3- to 3 (high)). This first sample we will
refer to as the $D \leq 60$~m sample, and contains 88 objects. We note
that the diameter measurement is an equivalent diameter computed from
the absolute magnitude and an assumed albedo. Such measurements
invariably contain some uncertainty but this is not quoted in the LCDB
and we do not discuss it here. For more information on the methods by
which these quantities are deduced the reader is directed to
\citet{warharpra09}.

It proves difficult to find derived axial ratios (or $a/b$ ratios) in
the literature, and most members of our first sample do not have
reported axis ratios.  So a second sample was selected to increase the
number of axis ratios available.  Since the LCDB doesn't quote the
necessary data, these asteroids are selected from the papers
referenced in paragraph 1 of this section. We had to increase the size
limit of the second sample to $\sim 300$m in order to obtain a sizable
sample of known axial ratios (92 asteroids in total).  We call this
second sample the $a/b$ ratio sample.

We note that asteroid shape -- specifically its $a/b$ ratio, assuming
a simplified triaxial ellipsoid shape where the axis lengths are $a
\leq b \leq c$ -- is not typically a parameter that is calculated in
most light curve studies. To overcome this, a formula presented by
\citet{kwibucodo10} was used in order to determine the minimum $a/b$
ratio from two parameters that are usually found in most surveys; the
light curve amplitude $A$ and the phase angle $\alpha$
(Eqn.~\ref{eq:abratio}).  We calculate the minimum axis ratio here
(that is, we assume equality in Eqn.~\ref{eq:abratio}) which thus is a
lower limit.

\begin{equation}
\frac{a}{b}\geq 10^{0.4 A(\alpha)/(1+0.03\alpha)} \label{eq:abratio}
\end{equation}

Non-principal axis (NPA) rotators (or tumbling asteroids) are also
taken into account here. NPA rotation is unstable rotation that occurs
when an asteroid is not spinning around its principal axis of maximum
inertia, a state which may be caused by an impact with a meteoroid or
another asteroid.  During NPA rotation, energy is slowly
dissipated from the asteroid until the body returns to stable
principal-axis rotation. Information on whether an asteroid was a
suspected tumbler is often recorded in the LCDB or the various surveys, though
it should be noted that these asteroids have not all been confirmed to
be tumblers. Some of these asteroids have been deemed possible
tumblers simply because of the irregularity of their light curves, and
further study is necessary to confirm NPA rotation. For the purposes
of this study, if an asteroid is either a confirmed or possible
tumbler, it has been designated as a tumbler in our samples. It is
noted in \citet{warharpra09} that there may be selection biases
against small fast-rotating tumblers due to the additional data
required to properly analyze a light curve with tumbling
characteristics, therefore there is a possibility that our study
underestimates the true fraction of tumblers in the general VSA
population.

\section{Results and Discussion}
 
\subsection{Rotation rate}
\begin{figure}[H]
\includegraphics[width=15cm]{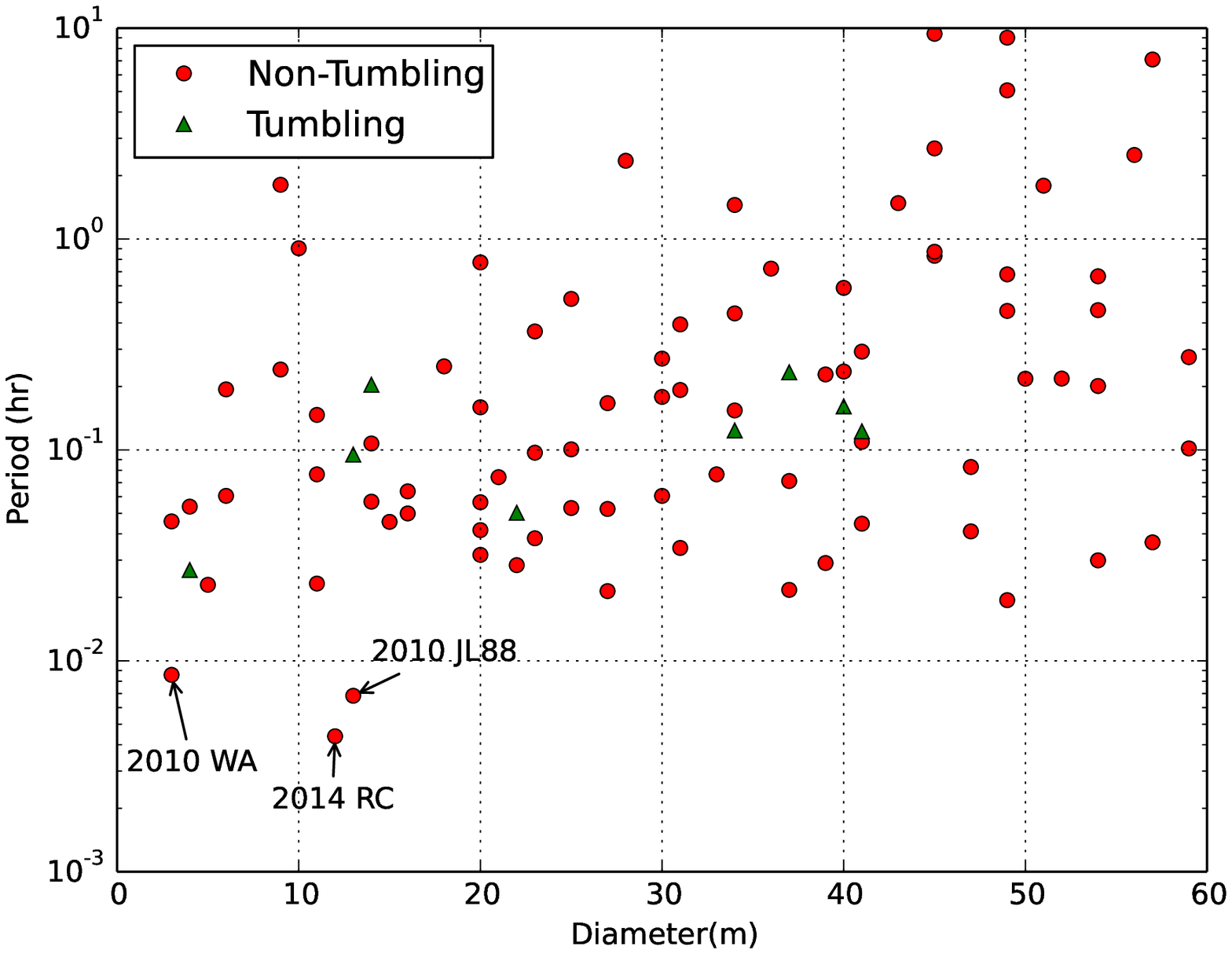}
\centering
\caption{Period versus diameter for the $D\leq 60$~m sample. Green triangles indicate known or suspected non-principal axis rotators.\label{Dvsp}. Asteroid 2014 RC is not part of this sample, but is added for reference.}
\end{figure}

Figure~\ref{Dvsp} is a plot of rotational period versus effective
diameter for our $D\leq 60$~m sample.  The typical fast rotational
nature of small asteroids mentioned in such papers as \citet{prahar00}
is apparent here with only 11 out of the 88 asteroids in the sample
having a period longer than one hour.  During the preparation of this
paper 2014~RC was discovered, with the fastest rotation period yet
reported at 15.8 seconds. We include it in Figure~\ref{Dvsp} for
reference though it has not been given a quality rating by the LCDB
yet and is technically not included in our $D\leq 60$~m
sample. Asteroid 2010~EX\textsubscript{11} (a 45~m diameter S-type)
has the slowest rotation period at 9.4 hrs. Overall, the mean period
was found to be 0.67 hrs or 40 minutes.

The two proposed fundamental types of structure for asteroids are
monolithic and ``rubble pile". Monolithic asteroids are made up of a
singular boulder, held together by its own tensile strength. On the
other hand, ``rubble pile" asteroids are made up from a collection of
gravitationally bound boulders, dust and regolith. These cannot spin
faster than what is commonly called the ``spin barrier"
\citep[e.g.,][]{harlar67,bur75, har96,prahar00} at around 2.2 hrs or
the asteroid will fly apart, though the shape of the asteroid will
affect the precise location of this boundary.  Our analysis here
confirms that a large portion of the VSA population consists of
fast-rotating asteroids supporting the suggestion of \citet{har96}
that they are monolithic.

All but three of the asteroids in Figure~\ref{Dvsp} have rotation
periods above 60 seconds, but we will make special note here of the
few that spin faster. The 3 meter S-type asteroid
2010~WA has a period of 31 seconds. 2010~JL\textsubscript{88} is a 13
meter diameter S-type which appears on the graph at 25 seconds. For
both of these asteroids the quality rating of the light curve data
collected is $U = 3$, which is the highest rating. We also note
asteroid 2014~RC, which has a diameter of 12 to 22 m and a rotation
period of 15.8
seconds\footnote{http://neo.jpl.nasa.gov/news/news185.html, Retrieved
  2014 Nov 9} though a full analysis of the observations has yet to be
published to our knowledge.

In its analysis of the asteroid capture process, the KISS report
considered the de-spin of a hypothetical asteroid with a period of 1
minute. Though this assumption is quite reasonable, it is worth noting
that there are a number of faster spinning asteroids in the current
observational sample.

We find that VSAs are likely to be rapidly rotating, and thus are perhaps
more likely to be held together by some tensile
strength, as opposed to a ``rubble pile".  \cite{hol03}
however notes that relatively small cohesive forces are needed to hold
a rubble pile together, far less than those present in dry terrestrial
soils.  It is also worth noting new observations made of a potential
ARM target 2011~MD, an S-type asteroid with a diameter of 7 meters and
a period of 11.6 minutes according to the LCDB.  New infrared scans from
NASA's Spitzer Space Telescope \citep{momfarhor14} indicate a
surprisingly high porosity, suggesting that it may be made up of a
collection of small boulders rather than being a singular
body. With its period being much faster than the spin barrier of 2.2
hrs, this implies that not all fast rotators are monolithic.

From Figure~\ref{Dvsp} it can be seen from the small number of green
triangles that the fraction of tumblers in the VSA population is
relatively small. Only eight out of the 88 asteroids collected in the
$D\leq60$~m sample have been deemed either confirmed or possible
tumblers in the LCDB. Four out of the 92 asteroids collected in the
$a/b$ ratio sample have been deemed either confirmed or possible
tumblers in the LCDB or the respective surveys in which their
observations were presented; none of these tumblers overlap with those
of the former sample. It would appear that tumblers constitute only a
very small part of the VSA population as a whole, but as
\cite{warharpra09} point out, the ratio of NPA rotators to principal
axis rotators may be greater than what is presented here due to an
inherent selection bias against small tumblers. We also note that the
ratio of tumbler to non-tumblers is likely to be a function of size,
axis ratios, etc.

\subsection{Axial ratio}

\begin{figure}[H]
\includegraphics[width=15cm]{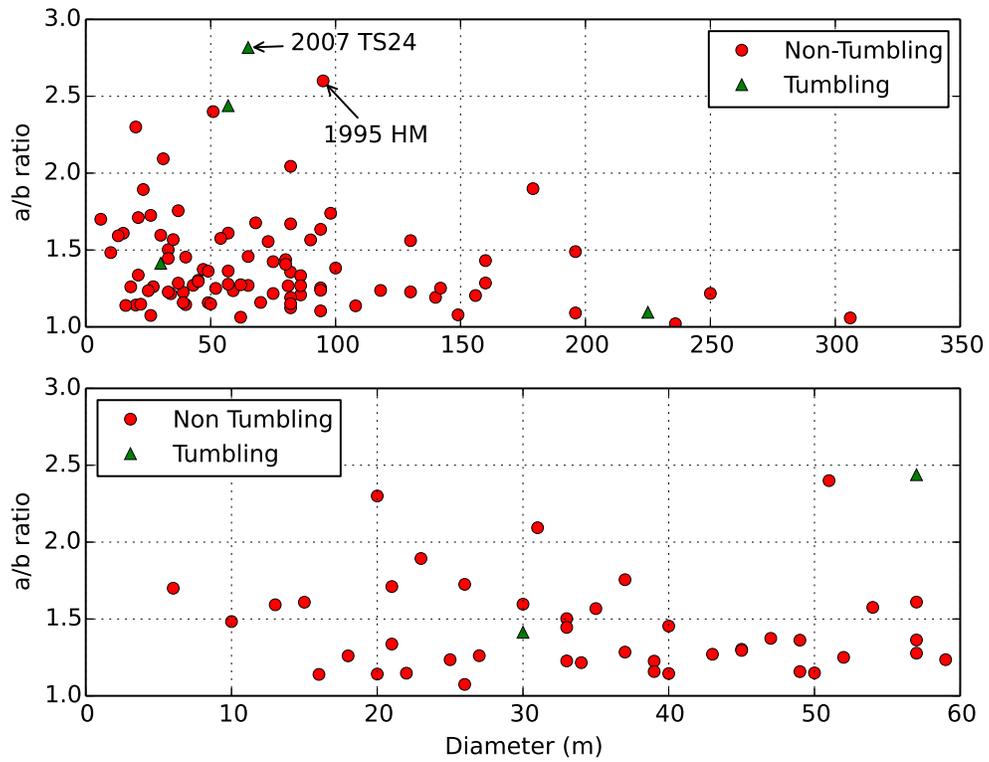}
\centering
\caption{The $a/b$ ratio versus diameter for the $a/b$ ratio sample (top), and a portion of the $a/b$ ratio sample with the restriction $D \leq$ 60m (bottom). \label{Dvsab}}
\end{figure}

A plot of axial ratio versus diameter is presented in
Fig~\ref{Dvsab}. The asteroid with the greatest axial ratio is
2007~TS\textsubscript{24} (a 65~m diameter S-type) at 2.8. This result
is derived directly in \citet{kwipolloa10}, although not without
some discussion that is worth noting. \citet{kwipolloa10} mention that the
strange light curve could be due to the asteroid being an NPA rotator,
and hence it is marked as such in the figures.

Another important result discussed in \citet{kwipolloa10} is with
respect to other asteroids with large $a/b$ ratios, specifically
1995~HM and 2000~EB\textsubscript{14}. Asteroid 1995~HM (a 94~m
diameter S-type) was originally analyzed in \citet{stemcngar97} and its unusual
light curve ascribed to a possible banana shape, but was
then re-analyzed in \citet{whithoher02} where it was given an
APR (amplitude-phase relation)-corrected axial ratio of 3.1, which
would give it the highest axial ratio known for VSAs. Asteroid
2000~EB\textsubscript{14} (a 51~m diameter S-type) was given an axial
ratio of 2.9 in \citet{whithoher02}, which would have placed it as the
second highest axial ratio.

\citet{kwipolloa10}, however, recomputed the results to be 2.6 and 2.4
for 1995~HM and 2000~EB\textsubscript{14} respectively, leaving 2007
TS\textsubscript{24} with the highest axial ratio, and 1995~HM with
the second highest. Since, as discussed earlier, we used the method of
$a/b$ ratio calculation of \citet{kwibucodo10}, we present their
result in the above graphs for consistency.  The precise value of
these axial ratios remains to be determined, but the important point
is that the current best upper limit for axial ratios with respect to
VSAs is around 3, and it would be unusual to find a VSA with an axial
ratio far above that. There are two caveats worth noting
however. Firstly, large axis ratios result in large magnitude
variations between telescopic exposures, and may cause high-amplitude
bodies to be missed entirely, biasing our sample. Secondly, our
method of determining axis ratios from the light curve from the formula
of \citet{kwibucodo10} provides a lower limit. As a result, the axis
ratios of small NEAs may be systematically larger than reported here.

\begin{figure}[H]
\includegraphics[width=15cm]{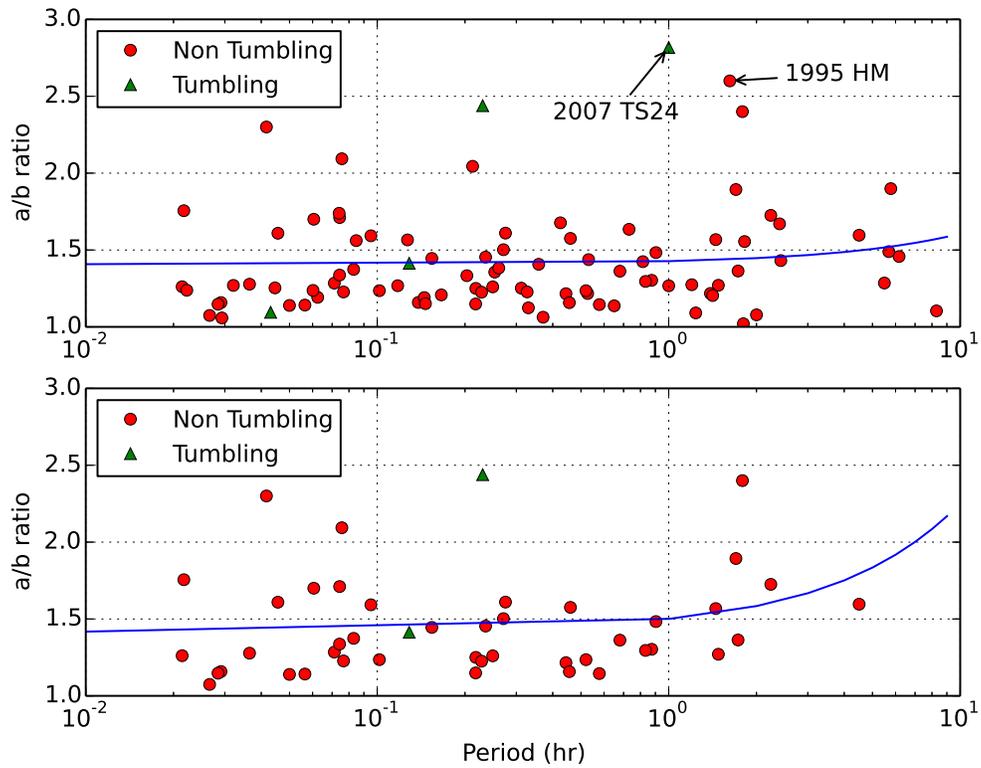}
\centering
\caption{The $a/b$ ratio versus rotation period for the full $a/b$ ratio sample (top) and the $D\leq60$~m restricted $a/b$ ratio sample (bottom). A least-squares linear fit to the data is presented in blue. The best fit line appears curved here because of the logarithmic x-axis.  \label{pvsab}}
\end{figure}

\citet{nakderyos11} concluded that small fast-rotating asteroids have a
tendency to be more spherical than slowly rotating asteroids, but
\citet{kwibucodo10} reported just the opposite. In Fig.~\ref{pvsab}
we find little correlation between the asteroid periods and their
$a/b$ ratios. Least squares fits to our samples do have slight
upslopes however, 0.0198 hr$^{-1}$ on the upper panel, and
0.0835 hr$^{-1}$ on the lower panel, so our samples do have a nominal
weak correlation. But these slopes are heavily leveraged by a few
points at the right-most edge of the figures and should be interpreted
with caution.

\begin{figure}[H]
\includegraphics[width=15cm]{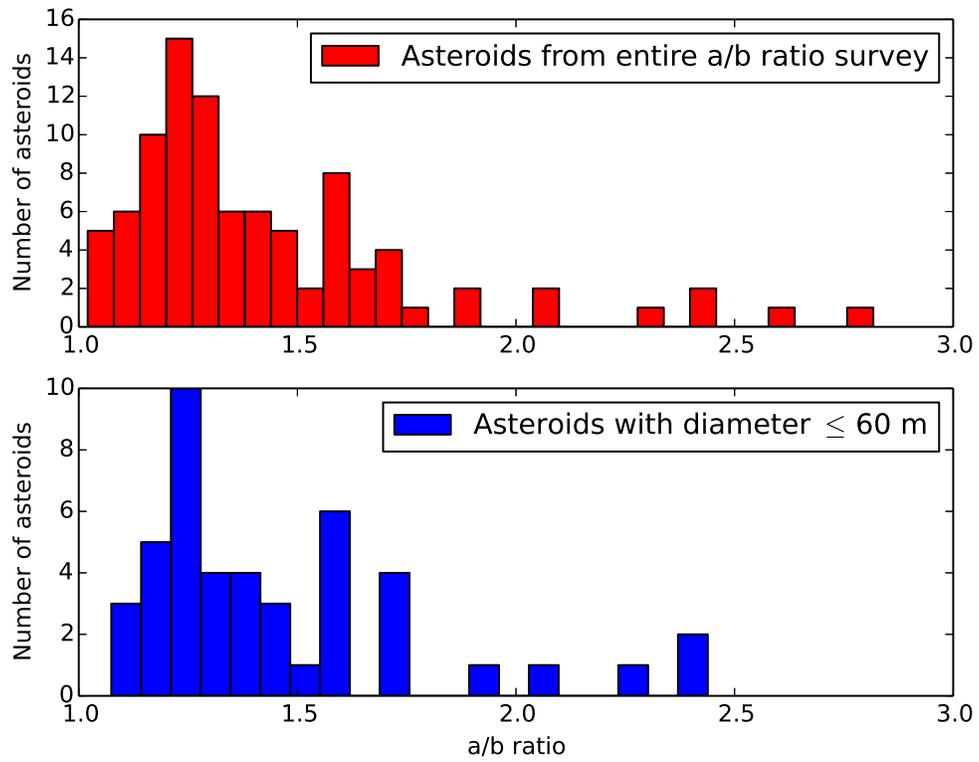}
\centering
\caption{Histograms of the axial ratios of the full $a/b$ ratio sample (top) and the $D\leq60$~m restricted $a/b$ ratio sample (bottom). \label{abhist}}
\end{figure}

Histograms of the axial ratios of our two samples are given in
Fig.~\ref{abhist}. The $a/b$ ratio sample has mean and median $a/b$
ratios of 1.43 and 1.29.  Our $D\leq60$~m sample does not have enough
information to compute axis ratios for all its members, but the
mean and median $a/b$ ratios of the 46 members of the $a/b$ sample
with diameters below 60~m are 1.46 and 1.36 respectively, consistent
with the idea that size and axis ratio are not strongly correlated.

We note that there has been some discussion in the literature
surrounding the determination of $a/b$ ratios in
\citet{nakderyos11}. Already in 2009, Warner et al. pointed out that
low light curve amplitudes (which result in concomitantly smaller axis
ratios) may simply be a result of finding the highest amplitude
spectral peak in noisy data.  In the LCDB itself, a significant
portion of the data are in the quality range of $U \leq1+$, meaning
that they are of doubtful quality. A fuller explanation as to why some
of this data were given such a low quality rating can be additionally
found in \citet{warharpra09}. 

\subsection{Taxonomic class and density}

In addition to period and effective diameter, the LCDB also records
the taxonomic class. Out of the 88 asteroids in our $D\leq60$~m
survey, 83 asteroids were of S-type (silicaceous ``stony" objects),
and out of the 92 asteroids in our $a/b$ ratio sample (which does
overlap partially with the previous sample), 89 were of S-type as
well, making it the most common type in our specific asteroid
population. The few other spectral types that were seen in the
population were four asteroids in the C-group (carbonaceous objects,
including one type B and one F), and three others being in the X-group
(metallic objects). It is believed that 20\% \citep{broculfri12} of
the near-Earth asteroid population is C-type, but that they are harder
to discover because of their lower albedos. Thus the C types are
underrepresented in our sample, reflecting the reality that our
observed sample is sharply limited by target brightness. We are not
arguing here that the real NEA population is low in C types, but the
set of potential targets for the ARM mission is likely to be.

Taxonomic class is linked to asteroid density, but for S-type
asteroids we must take the size into consideration as well since, as
\citet{car12} observes, the density of S-type asteroids appears to
increase with mass. If we look at Fig. 9 in the paper just mentioned,
S-type asteroids in the ARM size range would have an average bulk
density of around 2.6 g cm\textsuperscript{--3}, though the density of
S-type can be as low as 1.9 g cm\textsuperscript{--3} such as for
Itokawa \citep{fujkawyeo06}. Given the predominance of S types among
the small near-Earth asteroid population, it is reasonable to conclude
that the density of most potential ARM targets will be in the same
range though the density of a specifically chosen C-type target would be
lower, around 1 g~cm$^{-3}$ \citep{briyeohou02}. The composition, mass
and internal properties (rubble pile versus monolith) will all play a role here.

\section{Conclusions}

We have collected the available data on very small asteroids (VSAs)
with the highest quality light curves.  Unsurprisingly, a VSA will
most likely be found to have a period under the ``spin barrier" of 2.2
hrs; the average period from the $D \leq60$~m sample analyzed here is
0.67 hrs or 40 minutes. The lower limit for the period of the current
sample reaches down to 25 seconds (2010~JL\textsubscript{88}, a 13~m
diameter S-type) or even less (2014~RC, a 12-22 m Sq-type, with a
period of 16 seconds) though shorter periods are possible.

With respect to structure, our results imply that a VSA will probably
be a monolithic structure in which a singular boulder is held together by
its own tensile strength, as opposed to a ``rubble pile" in which
many boulders are gravitationally bound together, although arguments
from \cite{hol03} and new evidence from \citet{momfarhor14} show
that this may not necessarily be true.

We used the information on the light curves provided by various 
surveys to estimate the axial ratio. The VSAs in our samples have an
average minimum $a/b$ ratio of about 1.4, and the VSA with the
greatest axial ratio was found to be 2007~TS\textsubscript{24} at
2.8. Alternate analyses of some asteroid light curves have given
slightly different values, but all VSAs observed to date are
consistent with axial ratios less than three.  The mission outlined by
the KISS report discussed a capture bag capable of accommodating a
10x15 meter asteroid with a 2:1 axis ratio. Most ($> 90\%$) of our
$D\leq60$~m restricted sample have an $a/b$ ratio less than 2, but a
few exceed this value. We do note that our method of determining axial
ratios from \citet{kwibucodo10} provides the minimum axial ratio
consistent with the light curve amplitude, and so the values reported
here are lower limits.

The composition of most potential targets is likely to be rich in
silicates (S-type taxonomic class). The KISS study suggested that
C-type asteroids would make more interesting targets because of their
more diverse composition, which include water, carbon compounds, rock
and metal. However, such asteroids are not common within the currently
characterized small near-Earth asteroid population, though four
C-group (including sub-types B and F) and three X-types appear in our
sample. Though the real NEA population is not necessarily this low in
C types, the list of potential ARM targets is likely to be poorer in
carbonaceous bodies than might otherwise be expected.

\section{Acknowledgements} 
The authors thank Duncan Steel and David Asher for thoughtful comments
which much improved this paper. This work was supported in part by the
Natural Sciences and Engineering Council of Canada and the Canadian
Space Agency (CSA) through the ASTRO CSA Cluster.

\bibliographystyle{apalike}\biboptions{authoryear}
\bibliography{Wiegert}

\end{document}